\documentclass[
submission
,copyright
]{eptcs}

\usepackage[utf8]{inputenc}
\usepackage[T1]{fontenc}
\usepackage{amsmath,amssymb}
\usepackage{stmaryrd}
\usepackage{relsize}
\usepackage{graphicx}
\usepackage[final]{listings}
\usepackage{dafny}
\usepackage[T1]{fontenc}
\IfFileExists{luximono.sty}{\usepackage[scaled=0.81]{luximono}}{\usepackage[scaled=0.81]{beramono}}
\usepackage[svgnames]{xcolor}
\usepackage{url}
\usepackage{xspace}
\usepackage{pgfpages}
\usepackage{tikz}

\usetikzlibrary{calc,trees,positioning,arrows,chains,shapes.geometric,%
  decorations.pathreplacing,decorations.pathmorphing,shapes,%
  matrix,shapes.symbols}
\tikzset{
  >=stealth',
  chain/.style={
    rectangle,
    rounded corners = 6,
    fill=black!5,
    draw=black!20,
    minimum height=1.7em,
    text centered,
    on chain
  },
  line/.style={draw, very thick, <-},
  every join/.style={->, very thick, shorten >=1pt},
}

\let\oldsection=\section
\renewcommand{\section}[1]{\oldsection{#1}\setcounter{subsection}{-1}}
\let\oldsubsection=\subsection
\renewcommand{\subsection}[1]{\oldsubsection{#1}\setcounter{subsubsection}{-1}}

\makeatletter
\renewenvironment{enumerate}
{\begin{oldenumerate}%
    \setcounter{enumi}{\ifnum\@enumdepth=1 -1\else\arabic{enumi}\fi}%
  }
  {\end{oldenumerate}}
\makeatother
\let\oldmaketitle=\maketitle
\renewcommand{\maketitle}{\oldmaketitle%
  \setcounter{footnote}{-1}  
  \setcounter{section}{-1}
  \setcounter{table}{-1}
  \setcounter{figure}{-1}
  \setcounter{equation}{-1}
}

\renewenvironment{thebibliography}[1]{%
  \let\section=\oldsection  
  \begin{oldthebibliography}{#1}%
    \setcounter{enumiv}{-1}%
  }%
  {%
  \end{oldthebibliography}%
}
\newcounter{inlinelist}
\newcommand{\newinlinelist}{\setcounter{inlinelist}{-1}}
\newcommand{\inlinelistitem}{\refstepcounter{inlinelist}\arabic{inlinelist}) }

\newcommand{\figref}[1]{Fig.~\ref{#1}}     
\newcommand{\Figref}[1]{Figure~\ref{#1}}   

\providecommand{\event}{Formal-IDE 2014}  

\title{The {D}afny Integrated Development Environment}
\author{
  K. Rustan M. Leino
  \institute{Microsoft Research, Redmond, WA, USA}
  \email{leino@microsoft.com}
  \and
  Valentin W\"ustholz
  \institute{ETH Zurich, Department of Computer Science, Switzerland}
  \email{valentin.wuestholz@inf.ethz.ch}
}
\def\titlerunning{The Dafny IDE}
\def\authorrunning{K.R.M. Leino, V. W\"ustholz}

\begin{document}

\maketitle

\begin{abstract}
  In recent years, program verifiers and interactive theorem provers have become more
  powerful and more suitable for verifying large programs or proofs. This has demonstrated
  the need for improving the user experience of these tools to increase productivity and
  to make them more accessible to non-experts. This paper presents an integrated development
  environment for Dafny---a programming language, verifier, and proof assistant---that
  addresses issues present in most state-of-the-art verifiers: low responsiveness and lack
  of support for understanding non-obvious verification failures. The paper demonstrates several
  new features that move the state-of-the-art closer towards a verification environment that can provide
  verification feedback as the user types and can present more helpful information about the
  program or failed verifications in a demand-driven and unobtrusive way.
\end{abstract}

\section{Introduction}
\label{sect:Introduction}

Program verifiers and proof assistants integrate three major
subsystems.  At the foundation of the tool lies the logic it uses, for
example a Hoare-style program logic or a logic centered around type
theory.  On top of the logic sits some mechanism for automation, such as
a set of cooperating decision procedures or some proof search
strategies (e.g., programmable tactics).  The logic and automation
subsystems affect how a user interacts with the verification system,
as is directly evident in the tool's input language.  The third
subsystem is the tool's integrated development environment (IDE),
which in a variety of ways tries to reduce the effort required by the
user to understand and make use of the proof system.

In this paper, we present the IDE for the program verifier
Dafny~\cite{Leino2010,KoenigLeino:MOD2011}.  The IDE is an extension of Microsoft Visual
Studio (VS).  It goes beyond what has been done in previous IDEs (for
Dafny and other verification systems) in several substantial ways.
\begin{description}
\item[continuous processing]
  The IDE runs the program verifier in the background, thus providing
  \emph{design-time feedback}.  The user does not need to reach for a
  ``Verify now'' button.

  Design-time feedback is common in many tools.  For example, the
  spell checker in Microsoft Word is always on in this way.  Anyone
  who remembers from the 1980s having to invoke the spell checker
  explicitly knows what a difference this can make in how we think
  about the interaction with the tool; the burden of having to go
  through separate spelling sessions was transformed into the
  interaction process that is hardly noticeable.  Parsing and type
  checking in many programming-language IDEs is done this way, enabling
  completion and other kinds of IntelliSense context-sensitive
  editing and documentation assistance.
  The Spec\# verifier was the first to integrate design-time feedback
  for a verifier~\cite{Boogie:Architecture}.
  The jEdit editor for Isabelle~\cite{Wenzel:jEdit} also provides
  continuous processing in the background by running both a proof search
  and the Nitpick~\cite{Nitpick} checker which searches for
  counterexamples to the proof goal.

\item[non-linear editing]
  The text buffer can be edited anywhere, just like in usual
  programming-language editors.  Any change in the buffer will cause
  the verifier to reconsider proof obligations anywhere in the
  buffer.  (Since the Dafny language is insensitive to the order of
  declarations, the proof obligations that have to be reconsidered can
  occur both earlier and later in the buffer.)

  Although such non-linear editing seems obvious, it is worth noting
  that it is in stark contrast to common theorem prover IDEs like
  ProofGeneral \footnote{\url{http://proofgeneral.inf.ed.ac.uk}}
  and CoqIde \footnote{\url{http://coq.inria.fr}}, where the user manually moves a
  \emph{high water mark} in the buffer---anything preceding
  this mark in the buffer has already been processed by the system and
  is locked down to prevent editing,
  and anything following the mark has not been processed and can be
  freely edited.

\item[multi-threading]
  The Dafny IDE makes more aggressive and informed use of available
  multi-threaded hardware.  The number of concurrent threads used is
  adjusted dynamically, depending on what the verification tasks at
  hand are able to saturate.

  Although conceptually an obvious thing to do, the Dafny tool chain
  previously lacked the features to run separate verification tasks in
  parallel.  The use of multiple threads is especially noticeable when
  a file is just opened in the editor, since caches are cold at that
  time and everything needs to be verified.

  The Isabelle/jEdit editor \cite{Wenzel:jEdit,Sternagel2012} comes with support for multi-threading,
  which is motivated by the fact that it also supports non-linear editing and therefore
  offers more opportunities to parallelize verification tasks.
  The SPARK 2014 toolset~\cite{SPARK2014:toolset} also supports
  multi-threading, both in its translation from SPARK into the
  intermediate verification language Why3 and in the Why3 processing
  itself.

\item[dependency analysis and caching]
  The Dafny IDE caches verification results along with computed
  dependencies of what is being verified.  Before starting a new
  verification task, the system first consults the cache.  This
  feature makes the tool more responsive and reduces the user's wait
  times.

  Our users have found this to be the most useful of our features
  for making the interaction between user and system more effective.
  It is also what makes continuous processing desirable for large
  files.
  When a user gets stuck during a verification attempt, a typical
  response is to try many little input variations that might explain or
  remove the obstacle at hand.  It is during these times that the user
  needs the tool the most, so supporting fluid interactions at this
  time is of utmost importance.

  There has been a lot of work on caching, modifying, and replaying
  proofs for interactive proof assistants.  For proofs performed by
  SMT solvers, Grigore and Moskal worked on these things in the
  context of ESC/Java2 \cite{GrigoreMoskal:EditAndVerify}.

\item[showing information]
  Commonly, a verification system can supply various associated
  declarations automatically.  For example, common induction schemes
  may be constructed by default, some types and loop invariants may be
  inferred, and syntactic shorthands can reduce clutter in the program
  text.  Sometimes, a user may find it necessary to inspect this
  information.  The Dafny IDE attempts to make this information
  available via \emph{hover text}---when the user hovers the mouse cursor
  over a part of the program text, say, an identifier, any
  additional information about that identifier is displayed.  This
  makes the information easily accessible to users, but is at the same
  time not cluttering up the view of the program text.

  Note that in console-based interactive tools, for example like
  ACL2~\cite{ACL2:book}, the unobtrusive nature of information in
  hover text is difficult to achieve.  Such a tool has to either
  provide a set of commands that can be used to query information
  gathered by the tool or optimistically
  spill out a stream of information to the console window in
  the off-chance that a user wants to see some part of that information.

  An important consequence of making additional information easily
  accessible to the user is that it gives the verification system
  greater freedom in what can be computed automatically.  Users no
  longer need to fully understand the creative and elaborate schemes
  employed to compute this information, because whatever is computed
  can be viewed by the user, if needed.

  This feature is also common in programming-language IDEs, where
  inferred types or fully qualified identifier names are displayed as
  hover text.  The Dafny IDE takes this a step further, showing
  information such as default \emph{termination measures},
  specifications of implicit methods (such as those generated for
  \emph{iterators}), which calls are classified as
  \emph{co-recursive}, and code inherited by Dafny's ``\S{...}''
  construct from a refined module.

\item[integrated debugging]
  Verification error messages can have a lot of associated
  information, some of which can be useful to users.
  Previously, the Dafny IDE would highlight, directly in the IDE
  editor, the error trace leading to a reported error.  SPARK 2014
  also does this, for example.  To get information about the possible
  values of variables for the reported error, a Dafny user can use the Boogie
  Verification Debugger (BVD)~\cite{LeGouesLeinoMoskal7041}, which presents
  this information in a format akin to that provided in modern
  source-level debuggers.  We have done a deep integration of BVD into
  the Dafny IDE.

  Previously, BVD was accessible for Dafny only as a standalone tool,
  which meant the user manually had to correlate the source lines
  reported by BVD with the text buffer containing the program in the
  IDE.  The program verifier VCC~\cite{CohenDahlweidHillebrandLeinenbachMoskalSantenSchulteTobies2009} integrates BVD into
  its Visual Studio IDE.  The Dafny IDE now goes further, for example
  letting the user select which program state to inspect by clicking
  in the program text itself.  It also uses hover text to present
  values of variables in the selected state.
  OpenJML~\cite{OpenJML,OpenJML:F-IDE2014} also presents error information in this way,
  letting users inspect values of any subexpression and letting the
  source code location of the expressions hovered over determine which
  execution state is used to look up the value to be displayed.
\end{description}

As an alternative to running Dafny in Visual Studio,
Dafny can also be run from within a web browser (\url{http://rise4fun.com/dafny}) and from the
command line.  However, the bulk of the features we mention in this paper are
available only in the Visual Studio IDE extension.
Dafny, including its IDE, is available as
open source from \url{http://dafny.codeplex.com/}.

\section{Tool Architecture}
\label{sect:Tool_Architecture}

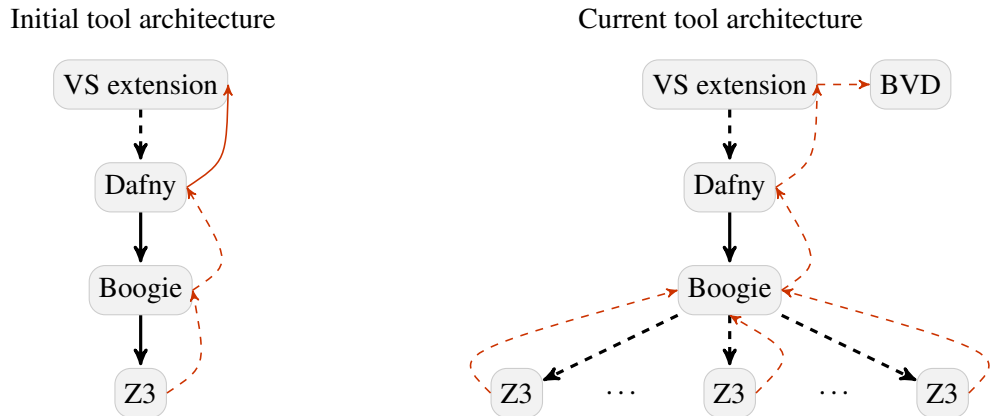
\begin{figure}[t]
  \centering
  \begin{minipage}[b]{0.45\linewidth}
    \centering
    Initial tool architecture

    \vspace{1em}

    \begin{tikzpicture}[
      node distance=.7cm,
      start chain=going below,
      ]
      \node[chain, join] (vse) {VS extension};
      \node[chain, join=by {very thick, dashed, ->}] (dafny) {Dafny};
      \node[chain, join] (boogie) {Boogie};
      \node[chain, join] (z3) {Z3};

      \draw[green!20!red, semithick, dashed, ->] (z3.east) .. controls ([yshift=10pt, xshift=15pt]z3.east) .. (boogie.east);
      \draw[green!20!red, semithick, dashed, ->] (boogie.east) .. controls ([yshift=10pt, xshift=15pt]boogie.east) .. (dafny.east);
      \draw[green!20!red, semithick, ->] (dafny.east) .. controls ([yshift=10pt, xshift=15pt]dafny.east) .. (vse.east);
    \end{tikzpicture}
  \end{minipage}
  \quad
  \begin{minipage}[b]{0.45\linewidth}
    \centering
    Current tool architecture

    \vspace{1em}

    \begin{tikzpicture}[
      node distance=.7cm,
      start chain=going below,
      ]
      \node[chain, join] (vse) {VS extension};
      \node[chain, join=by {very thick, dashed, ->}] (dafny) {Dafny};
      \node[chain, join] (boogie) {Boogie};
      \node[chain, join=by {very thick, dashed, ->}] (z3m) {Z3};
      \node[left = of z3m] (morel) {\ldots};
      \node[chain, left = of morel] (z3l) {Z3};
      \draw[very thick, dashed, ->] (boogie) -- (z3l);
      \node[right = of z3m] (morer) {\ldots};
      \node[chain, right = of morer] (z3r) {Z3};
      \draw[very thick, dashed, ->] (boogie) -- (z3r);

      \draw[green!20!red, semithick, dashed, ->] (z3l.west) .. controls ([yshift=15pt, xshift=-15pt]z3l.west) .. (boogie.west);
      \draw[green!20!red, semithick, dashed, ->] (z3m.east) .. controls ([yshift=15pt, xshift=15pt]z3m.east) .. (boogie.south);
      \draw[green!20!red, semithick, dashed, ->] (z3r.east) .. controls ([yshift=15pt, xshift=15pt]z3r.east) .. (boogie.east);
      \draw[green!20!red, semithick, dashed, ->] (boogie.east) .. controls ([yshift=10pt, xshift=15pt]boogie.east) .. (dafny.east);
      \draw[green!20!red, semithick, dashed, ->] (dafny.east) .. controls ([yshift=10pt, xshift=15pt]dafny.east) .. (vse.east);

      \node[chain, right = of vse] (bvd) {BVD};
      \draw[green!20!red, semithick, dashed, ->] (vse.east) -- (bvd.west);
    \end{tikzpicture}
  \end{minipage}
  \caption{Comparison of initial and current tool architecture. Arrows indicate data that
    is passed from one component to another, where dashed arrows indicate that data is
    transferred asynchronously. Less thick, red arrows indicate error information
    (including counterexamples for BVD in the current architecture) that is returned.}
  \vspace{-0.8em}
  \label{fig:Tool_Architecture}
\end{figure}

Before presenting the new tool architecture, we will give an overview of the
underlying components and the tool architecture that was used in the past (see
\figref{fig:Tool_Architecture}); it is similar to the architecture of other
verification tools that are built on top of the Boogie verification
engine~\cite{Boogie:Architecture}, such as Spec\#~\cite{BarnettFaehndrichLeinoMuellerSchulteVenter2011}
and VCC~\cite{CohenDahlweidHillebrandLeinenbachMoskalSantenSchulteTobies2009}. As the user
is editing the program, the VS extension continuously sends snapshots of the program to
the underlying Dafny verifier, which encodes the correctness proof
obligations as a translation into Boogie.
Boogie is an intermediate language~\cite{LeinoRuemmer:Boogie2} for program verification (similar to
Why3 \cite{FilliatrePaskevich2013}). Boogie programs typically consist of several
top-level declarations (e.g., axioms, variables, procedures) that are used to
formalize programs in a higher-level language, such as Dafny. For instance, each Dafny
\emph{method} is translated to
a Boogie procedure implementation that captures the well-definedness
conditions of the method's specification,
a Boogie procedure specification that captures the method
specification to be used by callers, and
a Boogie procedure implementation that captures the method body and
checks that it satisfies the method specification~\cite{Leino:Dafny:MOD2008}.
Similarly, each Dafny \emph{function}
is translated to a Boogie function and a Boogie procedure implementation that captures the
corresponding well-definedness conditions. The resulting Boogie program is sent to the
Boogie verifier, which generates verification conditions for each Boogie implementation to
discharge them using an automatic reasoning engine, typically the SMT-solver Z3
\cite{deMouraBjorner2008}. Verification errors that are revealed during this process are
propagated up to the VS extension, which displays them to the user.

This architecture gives rise to a pleasant and highly responsive user interaction for small
programs, but does not scale well to larger programs that consist of many methods and
functions. Since the requests to the underlying solver can easily be parallelized, we have
extended the Boogie verification engine to make use of separate tasks for verifying Boogie
implementations in parallel (using the \textsc{.Net} Task Parallel Library). Each task may
discharge its verification conditions using one or more solver instances that are managed
in a dynamically allocated pool of solvers. To take full advantage of this architectural
change, we made the propagation of verification errors to the user fully asynchronous (see
dashed arrows in \figref{fig:Tool_Architecture}).  This lets error messages show up as
soon as the corresponding verification condition has been processed by the
solver.  (Previously, Boogie only made use of multi-threading in one place,
namely in its mode for verification-condition
splitting~\cite{VC-splitting}.  We have preserved that functionality
and integrated it into the new task-based architecture.)

The Visual Studio extension for Dafny gets notified anytime there is a
new snapshot, that is, anytime the text buffer changes.  Upon each
such change, the extension recomputes syntax highlighting, which is
done through a simple lexical scan (that is, the parser is not invoked
and no abstract syntax tree is built).  After 0.5 seconds of
inactivity, the Dafny IDE invokes the Dafny parser, resolver, and type
checker on the current buffer snapshot.  If the snapshot passes these
phases without error, the additional information computed during these
phases (e.g., which calls are co-recursive) is made available to the
user in hover text.  Also, the snapshot is then asynchronously sent to
the Dafny verifier, unless the verifier is already running on a
previous snapshot.  As verification errors are reported by the
asynchronously running verifier, they are displayed in the IDE.
Once a snapshot has been fully processed by the verifier, a new verification task
is started for the current snapshot, unless that is the snapshot that was
just verified.

\begin{figure}
  \begin{minipage}[b]{0.32\linewidth}
    \centering
    \scalebox{0.74}{\includegraphics[trim={.9cm 25.85cm 42.9cm 2.85cm},clip]{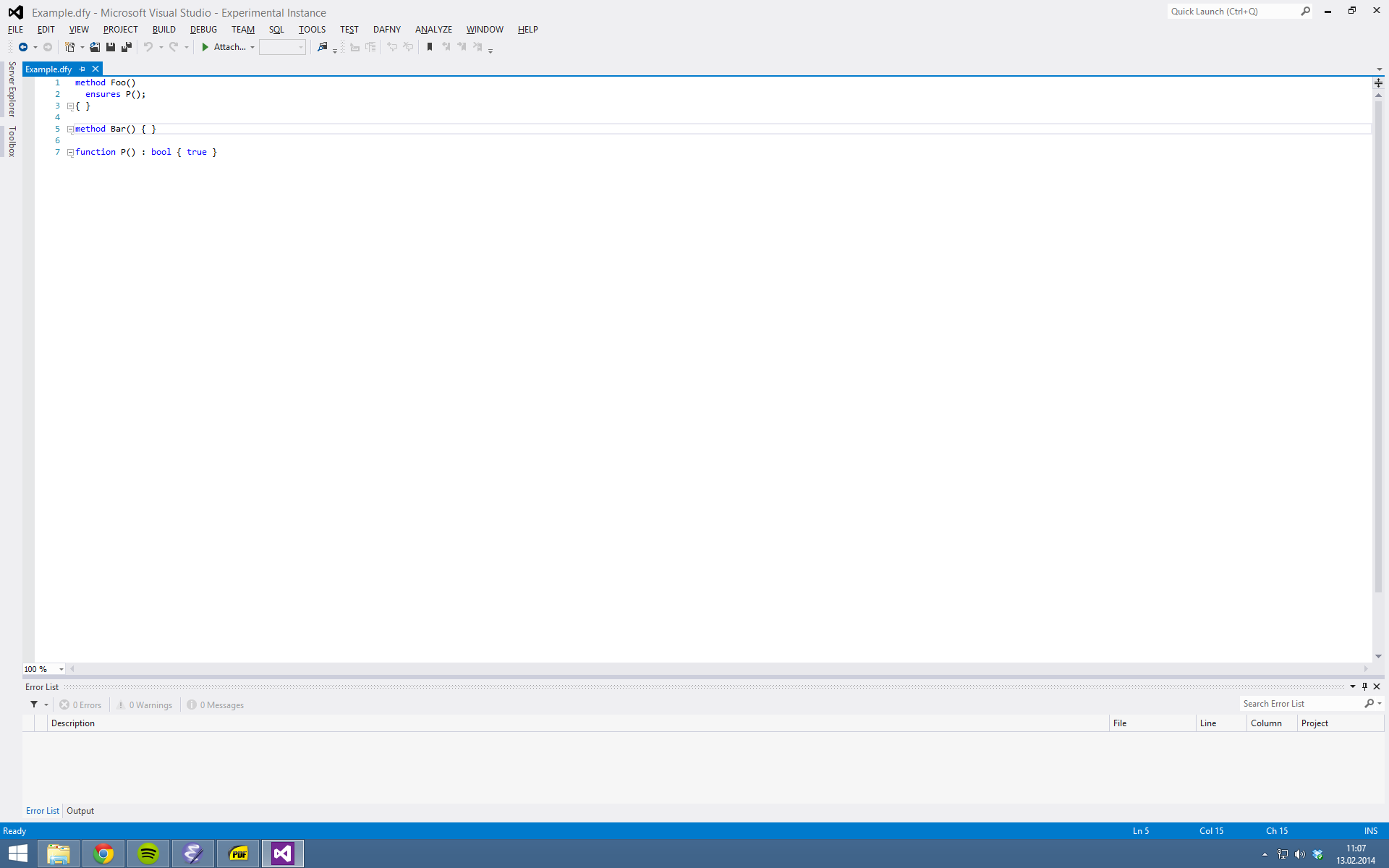}}
  \end{minipage}
  \quad
  \begin{minipage}[b]{0.32\linewidth}
    \centering
    \scalebox{0.74}{\includegraphics[trim={.9cm 25.85cm 42.9cm 2.85cm},clip]{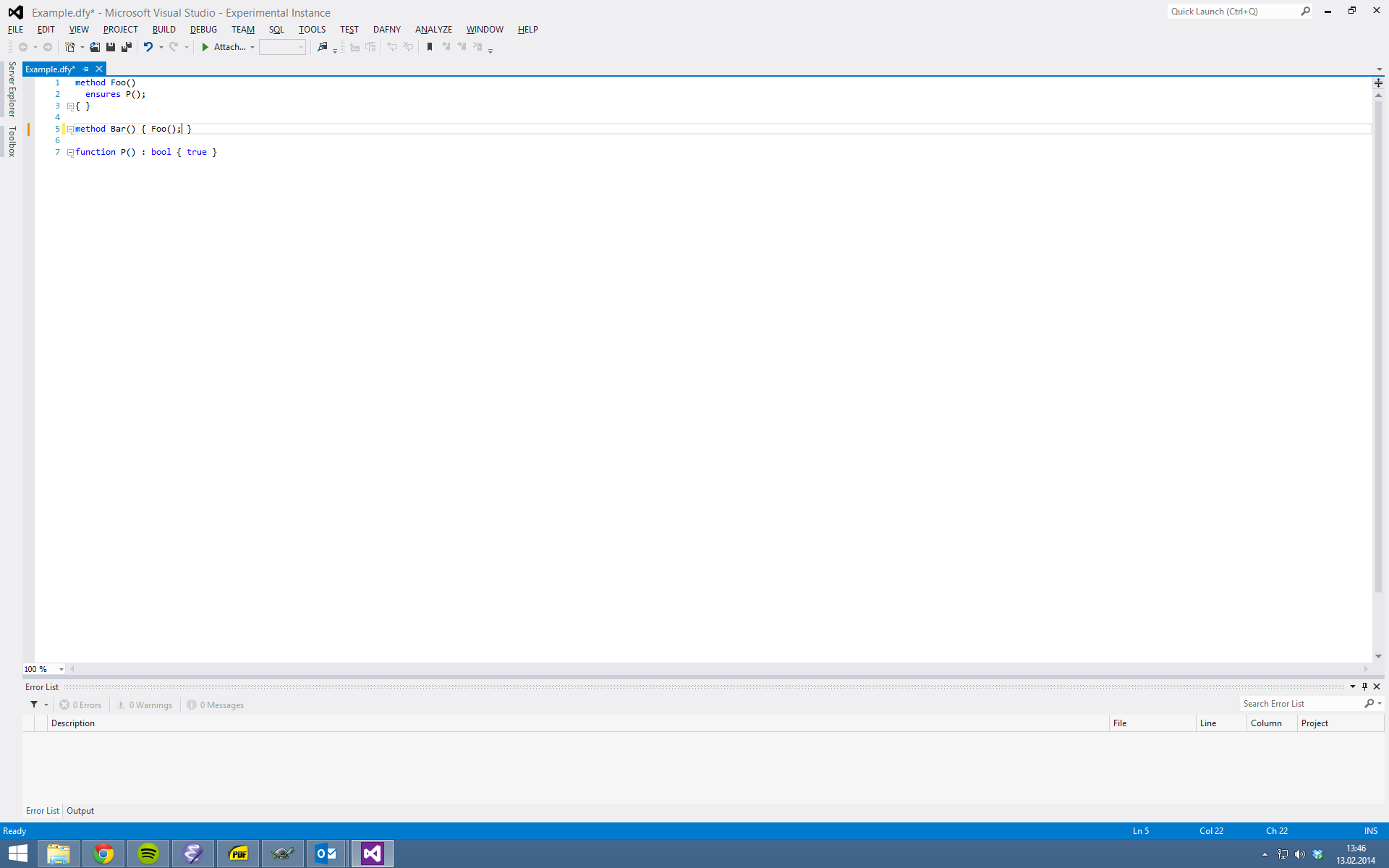}}
  \end{minipage}
  \quad
  \begin{minipage}[b]{0.31\linewidth}
    \centering
    \scalebox{0.74}{\includegraphics[trim={.9cm 25.85cm 42.9cm 2.85cm},clip]{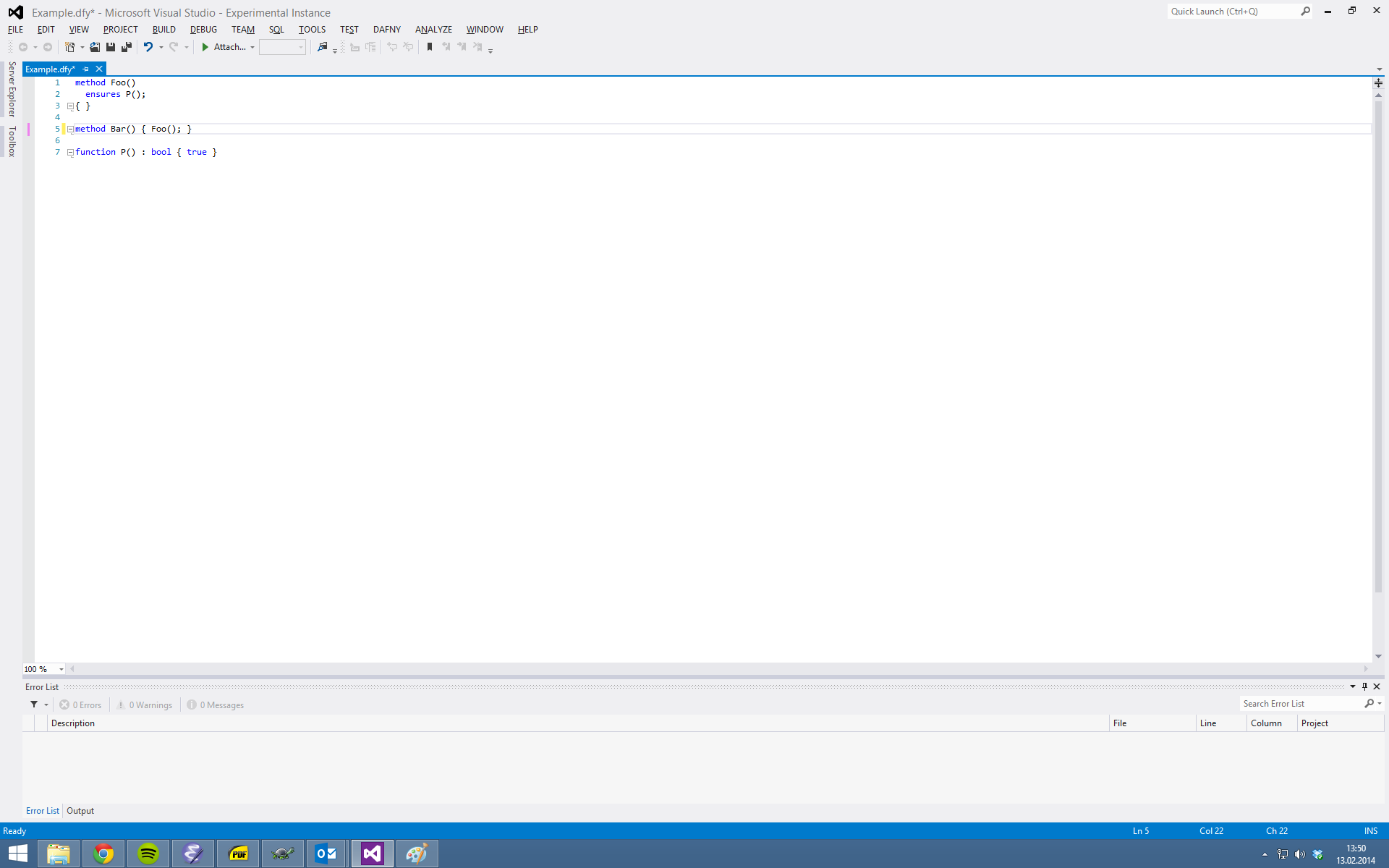}}
  \end{minipage}
  \caption{Progress indication via colors in the margins. The three program snapshots of
    the buffer are shown in chronological order (from left to right). The dark-orange
    margin in the middle snapshot indicates that changes have not yet been sent to the
    prover, while the purple margin in the right snapshot indicates that the verifier has
    started processing this snapshot.}

  \label{fig:screenshot:Margins}
\end{figure}

A constant question that users would have about Dafny's previous IDE was, ``Is the
verifier done yet?''.  To give the user a sense of the processing that is taking place in
the background, the new Dafny IDE uses colors in the margin (see Figure
\ref{fig:screenshot:Margins}).  A dark-orange color in the margin shows a line that has
been edited in a snapshot that has not yet been sent to the verifier, and a violet color
in the margin shows a line that has been edited in a snapshot that is currently being
processed by the verifier.

We also changed the tool architecture to integrate the Boogie
Verification Debugger (BVD) \cite{LeGouesLeinoMoskal7041} directly.
Under this change, which is independent of the parallelization, the
solver is asked to include the counterexample information needed by
BVD with each verification error.

\section{On-demand Re-verification}
\label{sect:On-demand_Re-verification}

Caching is a popular technique for improving the responsiveness of systems that
would need to repeatedly perform expensive computations whose output is a function of the given input.
Since in a modular verification approach different entities of a program (e.g.,
modules, classes, or---as in Dafny---methods and functions) are verified in isolation, changes to one
program entity usually invalidate only a small fraction of the verification results
previously obtained for other program entities. More specifically,
one can safely avoid re-verification of an entity by caching previously computed
verification results, except when the user has changed some other
program entity on which it depends.
This optimization is crucial in providing rapid feedback when the
program is larger than just a handful of entities.

Our technique for avoiding re-verification of methods and functions in Dafny deals with two
core issues:
\newinlinelist
\inlinelistitem detecting changes to program entities and
\inlinelistitem tracking dependencies between different program
entities to determine what needs to be re-verified.
To solve the first issue, we extended Dafny to compute an \emph{entity checksum} for each function, each
method, and the specification (e.g., pre- and postconditions) of each method. This checksum is
insensitive to various minor changes of the specific program text,
because it is computed based on the Dafny abstract
syntax tree.  For instance, the checksum of a method does not change if a comment is edited by
the user. To deal with the second issue, these entity checksums are used to track
dependencies by computing \emph{dependency checksums} for each program entity (function,
method, or method specification) based on its own entity checksum and the dependency checksums of
other entities on which it depends directly (e.g., methods it calls).
This lets us compare the
dependency checksum of a given entity for the current program snapshot with the one stored in
our verification result cache to determine if it needs to be re-verified.

In our implementation, we chose to compute the dependency checksums at
the level of Boogie entities, thus making this feature available to other verifiers
that target Boogie.  To set an entity checksum, a Boogie client (here,
Dafny) tags the entity with a particular \emph{custom attribute}---a
general mechanism supported by Boogie for attaching directives to
declarations---and gives the checksum as an integer argument to this
attribute.

Figure \ref{fig:On-demand_Re-verification_Example} illustrates how our technique works on a
concrete verification session that consists of three program snapshot, which are sent to
the prover in chronological order (i.e., snapshot 0 is the initial program and snapshot 2
is the final program). All entities of the initial program snapshot need to be verified,
since nothing has been cached yet. For snapshot 1, only method \texttt{Bar} needs to be
re-verified: the corresponding Boogie implementations (for checking the correctness and
well-definedness of the method body) are tagged with an entity checksum that is different
from the one in the cache, but the entity checksum of the corresponding Boogie procedure
(for capturing the method specification) stays the same. For snapshot 2, all entities need
to be re-verified: the entity checksum of the Boogie function that corresponds to the
Dafny function \texttt{P} changes with respect to the previous snapshot, which affects the
dependency checksums of all remaining Boogie implementations.

\begin{figure}[t]
  \centering
  \begin{minipage}[b]{0.31\linewidth}
    \centering
    Snapshot 0
    \begin{dafnyNarrowMargin}
method Foo()
  ensures P();
{ }

method Bar() { }

function P(): bool { true }
    \end{dafnyNarrowMargin}
  \end{minipage}
  \quad
  \begin{minipage}[b]{0.32\linewidth}
    \centering
    Snapshot 1
    \begin{dafnyNarrowMargin}
method Foo()
  ensures P();
{ }

method Bar() { ¤\underline{Foo();}¤ }  ¤\label{v1.changed_line}¤

function P(): bool { true }
    \end{dafnyNarrowMargin}
  \end{minipage}
  \quad
  \begin{minipage}[b]{0.31\linewidth}
    \centering
    Snapshot 2
    \begin{dafnyNarrowMargin}
method Foo()
  ensures P();
{ }

method Bar() { Foo(); }

function P(): bool { ¤\underline{\textbf{false}}¤ }  ¤\label{v2.changed_line}¤
    \end{dafnyNarrowMargin}
  \end{minipage}
  \caption{
    Example of on-demand re-verification. The three program snapshots are ordered
    chronologically (i.e., snapshot 0 is the initial program and snapshot 2 is the final
    program) and changes between snapshots are underlined. All entities in snapshot 0 need
    to be verified, while for snapshot 1 only method \texttt{Bar} needs to be
    re-verified. Finally, for snapshot 2 all entities need to be re-verified since all of
    them depend directly or indirectly on the modified function \texttt{P}.
  }
  \vspace{-1em}
  \label{fig:On-demand_Re-verification_Example}
\end{figure}

One interesting application of this technique has to do with prioritizing the program
entities that are being verified. Ideally, we want to prioritize entities that are more
directly affected by the latest change to the program text, because
that is where the user is likely to want to see the effect of the
re-verification first.  To do that, we assign different
levels of priority to an entity based on its current checksums and the ones stored in the
verification result cache:
\newinlinelist
\inlinelistitem low (current entity checksum is identical to the one in
the cache, but the dependency checksum is different; entity was unchanged, but some dependency
was changed),
\inlinelistitem medium (current entity checksum is different from the one in the cache; i.e.,
entity was changed directly),
\inlinelistitem high (no cache entry found; i.e., entity was added
recently), and
\inlinelistitem highest (current dependency checksum of the entity is identical
to the one in the cache). This prioritization scheme is motivated by the observation that users
usually prefer to get rapid feedback regarding the entities that were recently added or
changed directly. Note that we assign the highest priority to entities that were not
affected by the change at all, since displaying the corresponding verification results only
requires a simple cache lookup and we want to minimize the time during which the
corresponding errors are not displayed to the user. This prioritization scheme could
be extended easily to support more fine-grained priority levels.

Other verification systems have also used forms of checksums and
dependencies in order to reduce the need to construct new proofs.
In the heterogeneous Why3 system, both the construction and
verification of proof obligations can be parameterized by different
transformations and different solvers.  To maintain proofs as much as
possible when any subsystem changes, or if the program under scrutiny
changes, Why3 uses a scheme of \emph{proof sessions} and \emph{goal
shapes} for tracking
dependencies~\cite{BobotFilliatreMarcheMelquiondPaskevich2013}.  This
has let more than 100 program proofs be automatically maintained over
a period of more than two years.  For Dafny, we have focused on
reducing turnaround time for the user, rather than trying to be robust
against changes in components of Dafny itself.  Still, perhaps Dafny
could benefit from proof sessions and goal shapes as we, in the
future, move to tracking finer-grain dependencies.

Change management is also important in interactive proof assistants
where large parts of proofs are authored by users.  Work on such change
management has been done, for example, in the context of
KIV~\cite{DBLP:conf/fsttcs/ReifS93} and KeY~\cite{Klebanov:PhD}.

\section{User Interaction}
\label{sect:User_Interaction}

\subsection{Computed Information as Hover Text}

A verification system typically computes various properties that
determine how verification conditions are formulated.  For example,
Dafny uses heuristics to determine automatically generated induction
hypotheses~\cite{Leino:induction}.  Sometimes, it can be unclear to
the user which properties were computed.  For instance, Dafny uses some
rules that determine if a function self-call is recursive or
co-recursive; a user who does not know the precise rules may want to
find out which calls have been determined to be co-recursive.

We devised a simple mechanism by which the Dafny resolver and type
checker can associate any information with any AST node.  When Dafny
is running in the IDE, this information then gets displayed as
hover text for the region in the text buffer that corresponds to the respective AST
node. We use this mechanism to display the type and kinds of variables (e.g.,
``(ghost local variable) \S{x: List<<int>>}'' or
``(destructor) \S{List.head: T}''),
the default \S{decreases} clauses for methods and
functions~\cite{Leino2010},
the automatically generated conclusions of \S{forall} statements,
which methods are tail recursive,
which function calls are co-recursive,
the expansion of the syntactic sugar for calls to prefix predicates
and prefix methods~\cite{LeinoMoskal:Coinduction},
the class expansion of iterators, and code inherited from a refined
module through Dafny's ``\S{...}'' construct.

\subsection{Error Reporting}

When a verification attempt is not going through, a user has to debug
the cause.  For example, the executable program may be wrong, the
specifications may be wrong, the given proof of a lemma may be
incorrect, more information may be needed to make the proof go
through, or the problem could be caused by some incompleteness of the
SMT solver.

One way to debug such a situation is to ask the verifier questions
like ``does the following condition hold here?'' (which is done by
adding an \S{assert} statement in the program text) and ``can the
proof goal be met under this additional assumption?'' (which is done by
temporarily adding an \S{assume} statement in the program text).  This
kind of interactive dialog with the verifier is supported well in the
Dafny IDE, because the caching (and sometimes parallelization) makes
the interaction swift and fluid.

It is also possible to obtain more information about the failing
situation.  This is done by exploring the counterexample produced by
the solver.  The Boogie Verification Debugger (BVD), via a Dafny
plug-in, makes this counterexample intelligible at the source
context~\cite{LeGouesLeinoMoskal7041}.  BVD was previously available
for Dafny only as a standalone tool, but we have now integrated it
directly in the IDE.

\begin{figure}
  \scalebox{0.63}{\includegraphics[width=711pt]{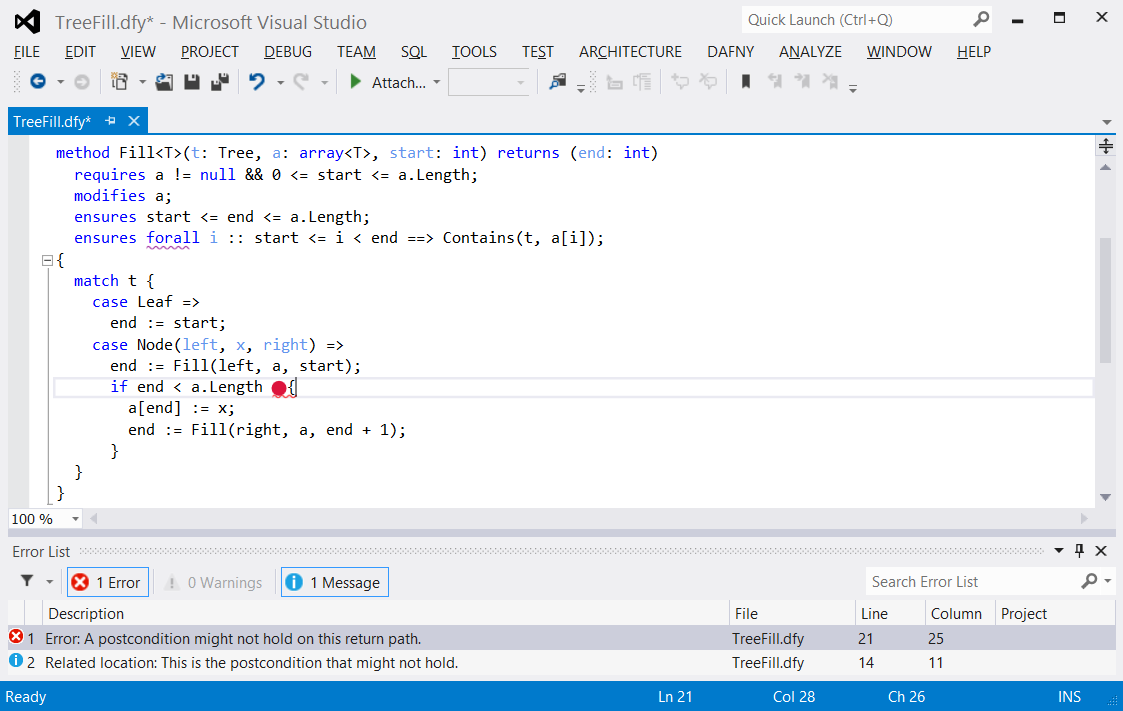}}
  \caption{A screenshot of the Dafny IDE.  The verification error is
    displayed in the text buffer as a red dot, which can be selected to
    obtain more information.}
  \label{fig:screenshot:RedDots}
  \vspace{3mm}
  \scalebox{0.63}{\includegraphics[width=711pt]{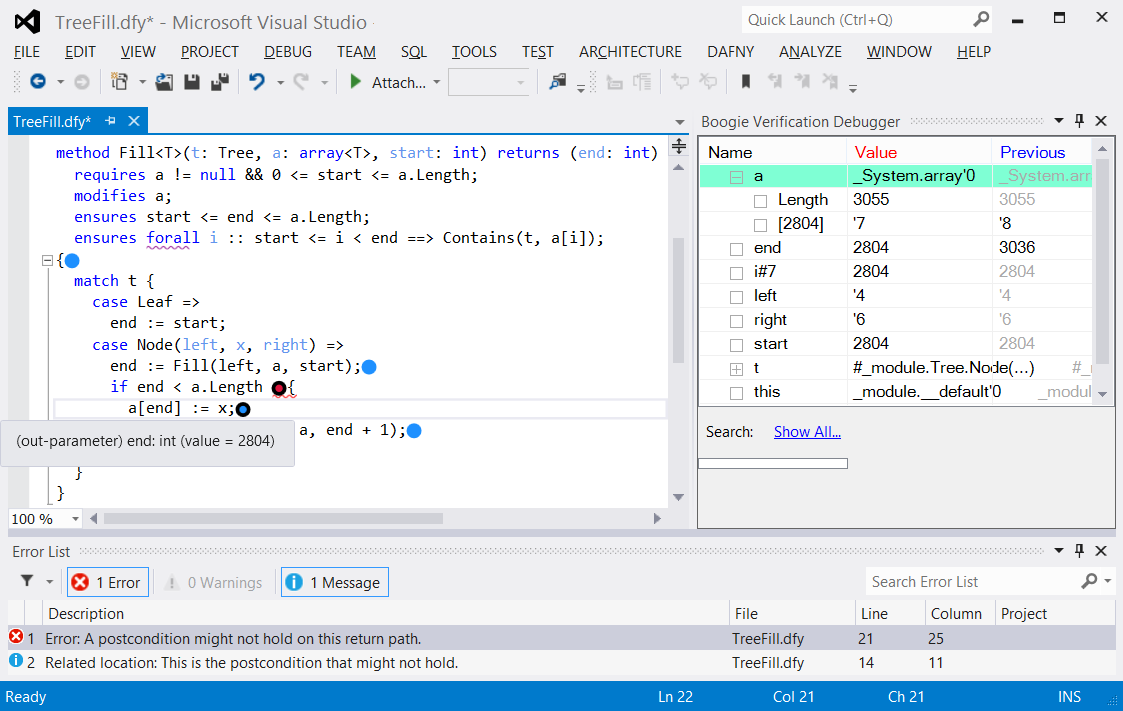}}
  \caption{A screenshot showing additional information obtained by
    selecting an error (red dot).  The blue dots show the program states
    along the control path leading to the error, and the BVD pane to the
    right shows values of variables in the selected state of the
    selected error.}
  \label{fig:screenshot:BlueDots}
\end{figure}

Let us describe our interface to BVD.  When an attempted verification
fails, like the postcondition violation shown in
\figref{fig:screenshot:RedDots}, a red dot (and a red squiggly
line) indicate the return path along which the error is
reported.  The error pane at the bottom of the screen shows the error
message, which also appears as hover text for the squiggly line.  The
error pane also lists source locations related to the error, in this
case showing the particular postcondition that could not be verified.

By clicking on a red dot, the Dafny IDE will display more information
related to that error, resulting in the screen shown in
\figref{fig:screenshot:BlueDots}.  The blue dots that now appear in
the program text trace the control path from the start of the
enclosing routine and leading to the error.  There is state
information associated with each blue dot and the user can click on a
blue dot to select a particular state (by default, the last state is
selected, which is the state in which the error was detected).

In addition to the blue dots, BVD is brought up in a pane to the
right.  BVD shows the variables in scope, in a familiar debugger-like
fashion, but with two conspicuous differences: some of the values
shown are underspecified (the names of these values begin with an
apostrophe, like \texttt{'7} and \texttt{'8}; distinct names refer to
distinct values), and some values are not shown at all, because they
are not relevant to the counterexample (like all of the array elements
of \S{a}, except the one at index \S{2804}).

The ``Value'' column in the BVD pane shows values in the currently
selected state, whereas the ``Previous'' column shows the values in
the previously selected state.  This gives a simple way to compare the
values in two states.  In the example in the figure, we had first had the error state
selected and then selected the state one line earlier.

Finally, the figure illustrates how values for variables of primitive types (in the
currently selected state) are also displayed as hover text.

What all of this tells us for the example is that the postcondition
cannot be verified when the bound variable \S{i} in the postcondition
is \S{2804}.  That index of the array is set by the assignment to
\S{a[end]}, but is then changed (from \texttt{'7} to \texttt{'8}) in
the next line where a recursive call to \S{Fill} is made.  Some
thinking then reveals that the cause of the verification error is that
the postcondition of \S{Fill} is too weak.  We can fix the problem by
adding a postcondition about the array indices between \S{0} and
\S{start}, in particular by saying that \S{Fill} leaves those array
elements unchanged:
\begin{dafny}
  ensures \forall i :: 0 <= i < start ==> a[i] == old(a[i]);
\end{dafny}
By simply typing in this extra postcondition and then waiting a split
second, the error goes away.

\section{Experience}
\label{sect:Experience}


\Figref{fig:Parallelization} shows some preliminary performance
numbers, comparing for some long-running verification tasks the effect
of using one solver versus three solvers. Two of the four programs
(\texttt{ParallelBuilds.dfy} and \texttt{LnSystemF.dfy}) were developed
by users of our tool, whereas the two other programs contain solutions to several verification
challenges from two different verification competitions.

\begin{figure}
  \[\begin{array}{lrrr}
    \mbox{Program} & \mbox{LOC} & \mbox{1 solver instance} & \mbox{3 solver instances} \\
    \hline
    \texttt{ParallelBuilds.dfy} &  881 &        572 &                269 \\
    \texttt{LnSystemF.dfy}      & 1736 &        354 &                109 \\
    \texttt{VSComp2010.dfy}     &  536 &         34 &                 26 \\
    \texttt{VSTTE2012.dfy}      & 1063 &        110 &                 71
  \end{array}\]
  \vspace{-1em}
  \caption{Preliminary performance numbers showing the effect of
    parallelization.  Times are in seconds.}
  \label{fig:Parallelization}
\end{figure}

As a proof that the cache is very important (actually, even more important than
parallelization) for enabling a highly responsive user interaction, we measured the
performance improvements gained by using the cache.  We considered 5
``versions'' of two long-running verification tasks.  The 5 versions
are 5 copies of the same program, but randomly changing one of the
checksums, as if a user had edited the program.  \Figref{fig:Caching}
gives the results, which show that using the cache allows the 5
versions to be verified in a total time that is a small increment over
verifying the program once. We did not perform these measurements for the two other
programs (\texttt{VSComp2010.dfy} and \texttt{VSTTE2012.dfy}), since they contain
collections of independent programs, which might not be a representative use case.

\begin{figure}
  \[\begin{array}{lrr}
    \mbox{Program} &
    \mbox{3 solver instances (5 runs)} & \mbox{3 solver instances (5 snapshots)} \\
    \hline
    \texttt{LnSystemF.dfy}      &  510\ (5 * 102)     &                  123 \\
    \texttt{ParallelBuilds.dfy} & 1345\ (5 * 269)     &                  311
  \end{array}\]
  \vspace{-1em}
  \caption{Comparison of verifying five versions of two programs with (third column)
    and without (second column) using the cache.  Times are in seconds.}
  \label{fig:Caching}
\end{figure}

The largest single project using Dafny is the Ironclad project in the
systems and security research groups at
Microsoft Research, which currently comprises about 30,000 lines of
Dafny specifications, code, and proofs.  The current Dafny IDE has
benefited from feedback from the Ironclad team.



\section{Conclusions and Future Work}
\label{sect:Conclusions_and_Future_Work}

The Dafny IDE represents a new generation of interaction between user
and verification system.  We have built dependency analysis,
caching, and concurrent verification into the design-time feedback
loop to make re-verification responsive with minimal user effort.  We
have provided a deeper integration of the Boogie Verification
Debugger, whereby it both displays information in the program text and
can be controlled directly from within the program text.  And using
hover text, we have given easy access to computed information without
cluttering up the user display.

The new IDE provides many significant improvements.  It has also let
us discover a number of areas where the user interaction can be
improved further.  We will mention a number of them here.

The most pressing problem is what to do with verification tasks that
require a long time.  At the moment, our IDE performs all verification
on a per-method (or per-function) basis.  When a method is long and
difficult, we often wish for breaking up the verification task into
smaller pieces.  Boogie has some facilities for
\emph{verification-condition (VC) splitting}~\cite{VC-splitting} and
\emph{selective checking}, but our Dafny IDE is currently not taking
advantage of these.  We would like to dynamically adjust the parameters
of VC splitting and selective checking based on previous verification
attempts, and we would like to fit this into a finer granularity of
caching.

An important special case is where the verifier runs out of time.
Subjectively, we find that time-outs occur in some part of any larger
proof attempt, especially those that involve large recursive
functions or non-linear arithmetic, \emph{while} the user is working
on getting the verification through.  That is, time-outs are often a
symptom of missing proof ingredients, and good performance tends to be
restored once the necessary ingredients have been supplied by the
user.  Time-outs during this time are bad, since they are on the
user's time.
We set the solver time-out to 10 seconds.  We do allow this default to
be overridden through Dafny custom attributes, but making it longer
rarely seems to help in situations where the verification attempt is
really missing information.  For a user to figure out what information
is missing (let alone which proof obligations are taking a long
time), the solver must end its proof search and return a
counterexample.  Currently, the verifier does not produce as much
information for verification attempts that time out as it does for
attempts that fail.  A more ambitious goal would be to try to
determine the cause of the time outs, perhaps by automatically trying
to analyze the solver logs that the Z3 Axiom Profiler gives access
to.  The Why3 system guards against time-outs by being able to run
several solvers at the same time~\cite{FilliatrePaskevich2013}.

There are also a number of places where we would like to improve the
Dafny plug-in for BVD.  For example, the current version does not let
users inspect values of functions in the counterexample.  We could
also imagine a special BVD mode targeted to illustrate the proof state
when Dafny is used to prove theorems (not verify programs).

Currently, all additional information that we display is computed
during Dafny's resolution and type checking phases.  There is also
some information that Dafny computes during the verification phase,
but our current machinery has no hooks for displaying this
information to the user.

While we hope to work on these items to further improve the Dafny IDE,
we hope that the current IDE will continue to be useful and that it
will inspire the IDEs of other verification systems.

\oldsubsection*{Acknowledgments}

We are grateful to Nada Amin and Maria Christakis for providing
benchmark programs, and to Micha{\l} Moskal for helping with the BVD
integration.  We also thank Maria for helpful comments on a draft of this paper
and Nada, Maria, Arjun Narayan, and Bryan Parno for feedback on the tool.

\bibliographystyle{eptcs}
\bibliography{dafny-ide}

\end{document}